\documentstyle[12pt,fullpage]{article}

\title{Sonoluminescence as instrument for the measurement of time parameters 
of fast photomultipliers}

\author{P.V.Vorob'ev, I.N.Nesterenko, A.R.Frolov, T.V.Shaftan}

\date{Institute of Nuclear Physics, 630090, Novosibirsk}

\begin{document}

\maketitle

\section{Abstract}

The apparatus description for control of the time parameters
of photomultipliers with high time resolution is described.
For generation of ultrashort light flashes have been used
sonoluminescence effect -- emission of the light flashes which is
appearing at the stressed interior of collapsing air microbubble by
sound wave in water.

\section{Introduction.}

The modern experiment demands of photoelements with a high time resolution.
That problems arises at diagnostic of short electron bunch by synhrotron
radiation and the high energy physics experiments, where is used often
the time of flight methods of measurements.
For this experiments it is necessary to realize metrological
attestation of photoelements. Therefore, it is very important to create
effective, inexpensive device for generation ultrashort light pulses.
This light source have to possess sufficient brightness and wide spectral
range. For generation of ultrashort light flashes have been used
the sonoluminescence (SL) effect.

The SL phenomenon was discovered in 1934 \cite{Frenz}
and until present time is remaining enigmatic one.
As it turned out, that SL flashes are possessing of unique parameters:
\begin{itemize}
\item[-] at the each flash emit $10^5 - 10^6$ of photons in spectral
range from 150 to 800 nm\cite{Barb};
\item[-] the emission spectrum of the flash fit to one of black body
at effective temperature about $50000^{\circ}K$;
\item[-] the flash duration is very small $\tau \le 10$ psec\cite{Moran,Barb};
\item[-] the dispertion of time interval between flashes is about 100 psec\cite{Moran};\\
\item[-] size of emission region is less than 3 $\mu m$;
\item[-] position stable of emission region is better than 3 $\mu m$
(at measurement by 1000 flashes) \cite{Moran}.
\end{itemize}

The good review about SL that we know and that we unknow one can be found
in \cite{Barb}.  Today amaizing temporal and spectral characteristics
of SL phenomena have not of satisfactory explanation. Neither the approach
of Shwinger \cite{Schwin} and his followers \cite{Liber,Eberl}, which is
basing of presentations about quantum vacuum in cavity with moving
acceleration wall, nor gasdynamics approach are not explaning observation
parameters of SL flashes. However, SL is very simply to observe.

\section{Apparatus for generating of sonolumimescence}

Many time was noted in \cite{Barb} the apparatus for observation of SL is
simple and inexpensive (fig.~\ref{1}) and its can be done in scholl
physical laboratory. The main part of the apparutus is spherical flask
with volume about 100 ml. This flask is filled up clean deaerated water
at temperature near $0^{\circ}C$.

The first spherical harmonic frequency of volumetric luquid oscillation is
equal:
$$ f = \frac{c}{R} $$
about 25 kHz and little bit one is shifted relatively calculated frequency
because of influence of flask wall. The oscillation is excited by four
piezoelectric transducers (PZT), which are glued at the flask equator.
The fifth PZT is placed to flask bottom and one is used only for control
of acoustic resonance. For excitation of four PZT is used highly stabilized
generator and powerful audioamplifier (100 W). The amplifier output is
loaded by consecutive oscillating circuit, made up by capacity of the PZT
and additional inductance.
This circuit is tuned to the frequency of basic mode of water-filled flask.
A standing spherical wave in water is producing a dynamical trap in
the flask center, in which may be captured gas bubble. This bubble is
produced by microboiler or quick dive in water thin glass stick.
The bubble behaviour depends on the amplitude of sound wave.
One can to emphasize four levels of excitation power at which bubble
behaviour qualitatively is changed:
\begin{itemize}
\item[1)] power level is sufficient for bubble keeping into the trap.
The bubble follow in flask center. However, the bubble life time is limited
and one quickly dissolve in deaerated water.
\item[2)] power level, at which the bubble is stable and one exists unrestrictly
long time.
\item[3)] "dancing bubble", at this power level bubble performs randomize
jumps with amplitude about of millimeter.
\item[4)] power level, at which "dance" is ceased and the bubble is stable.
In this time bubble brightness is changed. From begining this power level
SL is observed.
\item[5)] at greate power level bubble is destroyed.
\end{itemize}
The bubble existence and one stability is controled by fifth PZT drive
signal, which one may be observed as Lissague figure on oscilloscope screen.

\section{Time measurements}

The measurements time resolution of photomultiplier (PMT) was realized
with use of correlative method by Brown-Twiss's scheme (fig.~\ref{2}).
The apparatus includes two photomultipliers, which are placed at equal
distances from the flask center and opposite direction. The PMT's signals
have been amplified, formed by discriminators and 
get to input of time-to-digit converter (TDC). The signal of first PMT
get to start for TDC, and the impulse of other PMT after delay line stoped
of TDC. In this measurements is used broad band amplifiers (with band 2.5 GHz)
and discriminators with watching threshold.

The measurement results for pair of photomultipliers NCT-2 is presented
on fig.~\ref{3}. A value point of TDC is equal 20 psec (all scale 2000 psec).
The dispertion of time intervals distribution  is determined of time
resolution of this correlometer and one equaled 56 psec. The time resolution
of each PMT is less in $\sqrt{2}$ time and equal 40 psec.

\section{Conclusion}

The simple and inexpensive equipment, which is allowing to get light
flashes with a duration about 10 psec, following with a frequency
about 25 kHz have been made. In each flashes is emitted $10^5$ --- $10^6$
photons in spectral range from 200 to 800 nm.
The emission spectrum corresponds to radiation of black body with
temperature about 50000 K. The light source appears in practical like
point (size of emission region less 3 $\mu m$). For the measurement of
PMT time resolution have been used correlational method, i.e.
it is measured the distribution of time intervals between photocounts
from two PMT. The section pathband of registration electronics equal 2.5 GHz
and time resolution about 5 psec. For exception of the influence of the
PMT impulse amplitude to measurements results we use fast discriminators with 
watching threshold \cite{frolov}.
The time resolution of PMT with microchannel plate NCT-2 type, which is
developing by INP in collaboration with plant "Ekran" \cite{anash}, is 40 psec.
The dispertion of spread time of the electron avalanche, determining
of time resolution, was measured and one amount 40 psec at illumination
of all cathode.

Now we are considering the possibility of measurement of flash duration by 
other method, which based  on registration of visibility fluctuation
of interference image in Michelson's interferometer.

\newpage

\begin{figure}[ht]
\begin{picture}(50, 200)
\put(50,200){\special{em: graph art_s2.gif}}
\end{picture}
\vspace*{12mm}
\caption{Basic apparatus for generating sonoluminescence. LFG -- low
frequency generator, Amp -- amplifier, Osc -- oscilloscope, L -- inductivity.}
\label{1}
\end{figure}

\newpage

\begin{figure}[h]
\begin{picture}(50, 200)
\put(0,200){\special{em: graph art_s1.gif}}
\end{picture}
\vspace*{25mm}
\caption{The mounting scheme for measurement time parameters.
PMT1, PMT2 -- photomultipliers, A -- amplifiers, D -- discriminators,
DL -- delay line.}
\label{2}
\end{figure}

\newpage

\begin{figure}[ht]
\begin{picture}(50, 200)
\put(0,200){\special{em: graph  x_r_gra2.gif}}
\end{picture}
\caption{Measured distribution of the time intervals between photocounts.}
\label{3}
\end{figure}

\end{document}